\documentclass{article}
\usepackage{epsfig}
\usepackage{amsmath}
\usepackage{amsfonts}
\usepackage{amssymb}
\usepackage{euscript}
\usepackage{graphicx}
\usepackage{multicol}
\usepackage{lscape}
\usepackage{pstricks}
\usepackage{multirow}

\usepackage[cp1251]{inputenc}
\usepackage[russian]{babel}

\begin{document}
\sloppy
\begin{center}
{{\Large\bfseries The plasma mechanism of stabilization of neutron-excess nuclei \\}}
\end{center}

\begin{center}
{\Large

\itshape{ B.V.Vasiliev}
}
\end{center}

\section{ Neutron-excess nuclei and the neutronization}
\begin{figure}
\hspace{-3cm}
\includegraphics[scale=0.9]{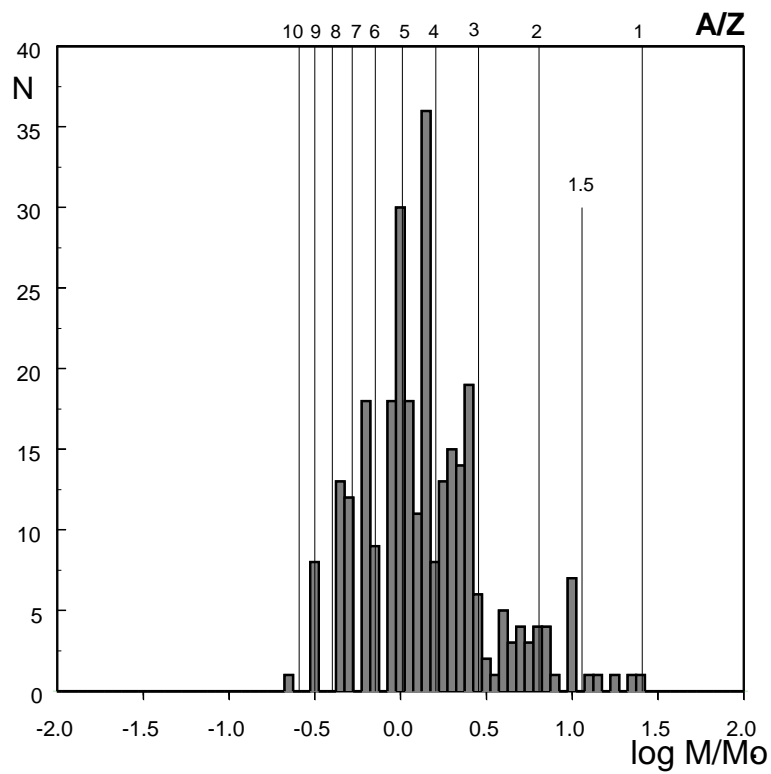}
\caption{The mass distribution of binary stars \cite{Heintz}. On
abscissa, the logarithm of the star mass over the Sun mass is shown.
Solid lines mark masses, which are calculated theoretically  \cite{BV}.}\label{starM}
\end{figure}

The form of the star mass spectrum (Fig.\ref{starM})
\cite{BV} indicates  that plasma of many stars consists of neutron-excess nuclei with the ratio $A/Z=3,4,5$ and so on. These nuclei are subjects of a decay under the "terrestrial" conditions.

Hydrogen isotopes  $_{ 1}^4H,~ _1^5H,~ _1^6H$ have short time of life and emit electrons with energy more than 20 Mev. The decay of helium isotopes $_2^6He,~ _2^8He, ~_2^{10}He$ have the times of life, which can reach tenth part of the seconds.

Stars have the time of life about billions years and the lines of their spectrum of masses are not smoothed. Thus we should suppose that there is  some mechanism of stabilization of  neutron-excess nuclei inside  stars. This mechanism is well known - it is neutronization \cite{LL}\S106. It is accepted to think that this mechanism is characteristic for dwarfs with density of particles about $10^{30}$ per cm$^3$ and
pressure of relativistic electron gas
\begin{equation}
{P}\approx\hbar c\cdot n_e^{4/3}\approx 10^{23}dyne/cm^2.\label{Pn}
\end{equation}

The possibility of realization of neutronization in dense plasma is considered below in detail.
At thus, we must try to find an explanation to characteristic features of the star mass spectrum.
At first, we can see that, there is actually quite a small number of stars with $A/Z=2$ exactly.
The question is arising: why there are so few stars, which are composed by very stable nuclei of helium-4? At the same time, there are many stars with $A/Z=4$, i.e. consisting apparently of a hydrogen-4, as well as stars with $A/Z=3/2$, which hypothetically could be composed by another isotope of helium - helium-3.

\subsection{The electron cloud of plasma cell}
Let us consider a possible mechanism of the action of the electron gas effect on the plasma nuclear subsystem.
It is accepted to consider  dense plasma to be  divided in plasma cells. These cells are filled by electron gas and they have positively charged nuclei in their centers \cite{Le}.

This construction is non stable from the point of view of the classical mechanics because  the opposite charges collapse is "thermodynamic favorable". To avoid a divergence in the theoretical description of this problem, one can  artificially  cut off the integrating at some small distance characterizing the particles interaction. For example, nuclei can be considered as  hard cores with the finite radii.

It is more correctly, to consider this structure as a quantum-mechanical object and to suppose that the electron can not approach the nucleus closer than its own de Broglie's radius $\lambda_e$.

Let us consider the behavior of the electron gas inside the plasma cell. If to express the number of electrons in the volume $V$ through the density of electron $n_e$, then the maximum value of electron momentum \cite{LL}:
\begin{equation}
p_F=\left(3\pi^2~n_e\right)^{1/3}\hbar.\label{pFn}
\end{equation}

The kinetic energy of the electron gas can be founded from the general expression
for the energy of the Fermi-particles, which fills the volume $V$ \cite{LL}:
\begin{equation}
\mathcal{E}=\frac{Vc}{\pi^2 \hbar^3}\int_0^{p_F}p^2\sqrt{m_e^2
c^2+p^2}dp.
\end{equation}
After the integrating of this expression and the subtracting of the energy at rest, we can calculate the kinetic energy of the electron:
\begin{equation}
\mathcal{E}_{kin}=\frac{3}{8}m_e c^2
\left[\frac{\xi(2\xi^2+1)\sqrt{\xi^2+1}-Arcsinh(\xi)-\frac{8}{3}\xi^3}{\xi^3}\right]\label{ekk}
\end{equation}
(where $\xi=\frac{p_F}{m_ec}$).

The potential energy of an electron is determined by the value of the attached electric field.
The electrostatic potential of this field $\varphi(r)$ must be equal to zero at infinity.\footnote{In general, if there is an uncompensated electric charge inside the cell, then we would have to include it to the potential
$\varphi (r)$. However, we can do not it, because will consider only electro-neutral cell, in which the charge of the nucleus exactly offset by the electronic charge, so  the electric potential on cell border is equal to zero.}
With this in mind, we can write the energy balance equation of electron
\begin{equation}
\mathcal{E}_{kin}=e\varphi(r).\label{ek-p}
\end{equation}
	
The potential energy of an electron at its moving in an electric field of the nucleus  can be evaluated on the basis of the Lorentz transformation \cite{LL2}\S24.		
If in the laboratory frame of reference, where an electric charge placed, it creates an electric potential $\varphi_0$, the potential in the frame of reference moving with velocity $v$ is
\begin{equation}
\varphi=\frac{\varphi_0}{\sqrt{1-\frac{v^2}{c^2}}}.
\end{equation}
Therefore, the potential energy of the electron in the field of the nucleus can be written as:
\begin{equation}
\mathcal{E}_{pot}= -\frac{Ze^2}{r}\frac{\xi}{\beta}\label{epp}.
\end{equation}
Where
\begin{equation}
\beta=\frac{v}{c}.
\end{equation}
and
\begin{equation}
\xi\equiv\frac{p}{m_e c},
\end{equation}
$m_e$ is the mass of electron in the rest.

And one can  rewrite the energy balance Eq.(\ref{ek-p}) as follows:
\begin{equation}
\frac{3}{8}m_ec^2\xi\mathbb{Y}=e\varphi(r)\frac{\xi}{\beta}
\label{ek-p2}.
\end{equation}
where
\begin{equation}
\mathbb{Y}=\left[\frac{\xi(2\xi^2+1)\sqrt{\xi^2+1}-Arcsinh(\xi)-\frac{8}{3}\xi^3}{\xi^4}\right].
\end{equation}
Hence
\begin{equation}
\varphi(r)=\frac{3}{8}\frac{m_ec^2}{e}\beta\mathbb{Y}.\label{fy}
\end{equation}

In according with Poisson's electrostatic equation
\begin{equation}
\Delta\varphi(r)=4\pi e n_e
\end{equation}
or at taking into account that the
electron density is depending on momentum (Eq.(\ref{pFn})),
we obtain
\begin{equation}
\Delta\varphi(r)=\frac{4e}{3\pi}\left(\frac{\xi}
{\lambda_C\hspace{-0.35cm}\widetilde{}}{~~}\right)^3,
\end{equation}
where
$\lambda_C{\hspace{-0.35cm}\widetilde{}}{~~}=\frac{\hbar}{m_e c}$
is the Compton radius.
\vspace{1cm}

At introducing of the new variable
\begin{equation}
\varphi(r)=\frac{\chi(r)}{r},
\end{equation}
we can 	transform the Laplacian:
\begin{equation}
\Delta\varphi(r)=\frac{1}{r}\frac{d^2\chi(r)}{dr^2}.
\end{equation}
As (Eq.(\ref{fy}))
\begin{equation}
\chi(r)=\frac{3}{8}\frac{m_ec^2}{e}\mathbb{Y}\beta r~,
\end{equation}
the differential equation can be rewritten:
\begin{equation}
\frac{d^2\chi(r)}{dr^2}=\frac{\chi(r)}{\mathbb{L}^2}~,\label{dL}
\end{equation}
where
\begin{equation}
\mathbb{L}=\left(\frac{9\pi}{32}\frac{\mathbb{Y}\beta}
{\alpha\xi^3}\right)^{1/2}
\lambda_C\hspace{-0.35cm}\widetilde{}~~~,
\end{equation}
$\alpha=\frac{1}{137}$ is the fine structure constant.
		
This differential equation has the solution:
\begin{equation}
\chi(r)= C\cdot exp\left(-\frac{r}{\mathbb{L}}\right).
\end{equation}
Thus, the equation of equilibrium of the electron gas inside a cell (Eq.(\ref{ek-p2})) obtains the form:
\begin{equation}
\frac{Ze}{ r}\cdot e^{-r/\mathbb{L}}=\frac{3}{8}m_ec^2\beta\mathbb{Y}~~.\label{epp3}
\end{equation}

\section{The Thomas-Fermi screening}
Let us consider the case when an ion is placed at the center of a cell, the external shells don't permit  the plasma electron to approach to the nucleus on the distances much smaller than the Bohr radius.
The electron moving is non-relativistic in this case.
At that $\xi\rightarrow 0$,
the kinetic energy of the electron
\begin{equation}
\mathcal{E}_{kin}=\frac{3}{8}m_ec^2\xi\mathbb{Y}\rightarrow \frac{3}{5}E_F~~,
\end{equation}
and the screening length
\begin{equation}
\mathbb{L}\rightarrow \sqrt{\frac{\mathcal{E}_F}{6\pi e^2 n_e}}.
\end{equation}
	Thus, we get the Thomas-Fermi screening in the case of the non-relativistic motion of an electron.
\section{The screening with relativistic electrons}
In the case the $\ll $bare$ \gg $ nucleus, there is nothing to prevent the electron to approach it at an extremely small distance $\lambda_{min}$, which is limited by its own than its de Broglie's wavelength. Its movement in this case becomes relativistic at $\beta\rightarrow 1$ и $\xi\gg 1$.
In this case, at not too small $\xi$, we obtain
\begin{equation}
\mathbb{Y}\approx 2\left(1-\frac{4}{3\xi}\right),
\end{equation}
and at $\xi\gg 1$
\begin{equation}
\mathbb{Y}\rightarrow 2 ~.
\end{equation}

In connection with it, at the distance $r\rightarrow \lambda_{min}$ from a nucleus, the equilibrium equation  (\ref{epp3})
reforms to
\begin{equation}
\lambda_{min}\simeq Z\alpha \lambda_C~~.
\end{equation}
and the density of electron gas in a layer of thickness $\lambda_{min}$ can be determined from the condition of normalization.
As there are Z electrons into each cell, so
\begin{equation}
Z\simeq n_e^{\lambda}\cdot{\lambda_{min}}^3
\end{equation}
From this condition it follows that
\begin{equation}
\xi_{\lambda}\simeq \frac{1}{\alpha Z^{2/3}}\label{xil}
\end{equation}
Where $n_e^\lambda$ and $\xi_{\lambda}$ are the density of electron gas and the relative momentum of electrons at the distance $\lambda_{min}$ from the nucleus.
In accordance with Eq.({\ref{ekk}}), the energy of all the Z electrons in the plasma cell is
\begin{equation}
\mathcal{E}\simeq Zm_ec^2 \xi_\lambda
\end{equation}
At substituting of Eq.({\ref{xil}}), finally we obtain the energy of the electron gas in a plasma cell:
\begin{equation}
\mathcal{E}\simeq\frac{m_e c^2}{\alpha} Z^{1/3}\label{z13}
\end{equation}
This layer provides the pressure on the nucleus:
\begin{equation}
{P}\simeq \mathcal{E}\left(\frac{\xi}{\lambda_C\hspace{-0.35cm}
\widetilde{}}{~~}\right)^3\approx 10^{23}dyne/cm^2
\end{equation}
This pressure is in order of value with pressure of neutronization (\ref{Pn}).
\section{The neutronization}.
The considered above $\ll$ attachment$\gg$ of the electron to the nucleus in a dense plasma should lead to a phenomenon neutronization of the nucleus, if it is energetically favorable. The $\ll$ attached $\gg$ electron  layer should have a stabilizing effect on the neutron-excess nuclei, i.e. the neutron-excess nucleus, which is instable into substance with the atomic structure,  will became stable inside the dense plasma. It explains the stable existence of stars with a large ratio of A/Z.

These formulas allow to answer questions related to the characteristics of the star mass distribution.
The numerical evaluation of energy the electron gas in a plasma cell gives:
\begin{equation}
\mathcal{E}\simeq\frac{m_e c^2}{\alpha} Z^{1/3}\approx 1.1 \cdot 10^{-4}Z^{1/3} erg
\end{equation}
The mass of nucleus of helium-4 $M(_2^4He)=4.0026~ a.e.m.$, and the mass hydrogen-4 $M(_1^4H)=4.0278~ a.e.m.$. The mass defect
$\approx 3.8\cdot 10^{-5} egr$.
Therefore, the  reaction
\begin{equation}
_2^4He+e\rightarrow _1^4 H,
\end{equation}
is energetically favorable. At this reaction the nucleus captures the electron from gas  and proton  becomes a neutron.
It explains why there are a small number of stars with $A/Z=2$ - only stars consisting of deuterium  are stable here.

There is the visible line of stars with the ratio $A/Z=3/2$ at the star mass spectrum. It can be attributed to the stars, consisting of $_2^3He$,~$_4^6Be$,~$_6^9C$, etc.

It is not difficult to verify at direct calculation that the reactions of neutronization  and transforming of $_2^3He$ into $_1^3H$ and $_4^6Be$ into $_3^6Li$ are energetically allowed. So the nuclei $_2^3He$ and $_4^6Be$ should be  conversed by neutronization into  $_1^3H$ and $_3^6Li$. The line $A/Z=3/2$ of the star mass spectrum  can not be formed by these nuclei.
However, the reaction
\begin{equation}
_6^9 C+e\rightarrow _5^9B~,
\end{equation}
is not energetically allowed
and therefore it is possible to believe that the stars of the above line mass spectrum may consist of carbon-9.

\vspace{1.6cm}

It is interesting, is it possible to study this effect under laboratory conditions?
One can use gaseous tritium for it  and can measure the rate of its  decay at two states.
First the rate of tritium decay  at atomic (molecular) state. Second, the same rate  under the effect of ionizing electric discharge. Essentially, one must aim to use the dense gas and the power discharge
in order to ionize more atoms of tritium. Taking into account the preceding consideration, such a test does not
look as a hopeless project.

\end{document}